\newcommand{\myname}{Geoff Boeing}
\newcommand{\myemail}{boeing@usc.edu}
\newcommand{\myaffiliation}{Department of Urban Planning and Spatial Analysis\\Sol Price School of Public Policy\\University of Southern California}

\newcommand{\papertitle}{Online Rental Housing Market Representation and the Digital Reproduction of Urban Inequality}
\newcommand{\papercitation}{Boeing, G. 2019. \papertitle. \textit{Environment and Planning A: Economy and Space}. \href{https://doi.org/10.1177/0308518X19869678}{https://doi.org/10.1177/0308518X19869678}}
\newcommand{\paperkeywords}{craigslist, demographics, digital divide, critical GIS, housing market, housing search, inequality, rental housing, smart cities, spatial analysis, spatial econometrics}

\RequirePackage[l2tabu,orthodox]{nag}
\documentclass[11pt,letterpaper]{article}
\usepackage[T1]{fontenc}
\usepackage[utf8]{inputenc}
\usepackage{crimson}
\usepackage{helvet}
\usepackage[strict,autostyle]{csquotes}
\usepackage[USenglish]{babel}
\usepackage{microtype}
\usepackage{abstract}                   
\usepackage{authblk}                    
\usepackage{booktabs}                   
\usepackage{caption}                    
\usepackage[final]{draftwatermark}      
\usepackage{endnotes}                   
\usepackage{geometry}                   
\usepackage{graphicx}                   
\usepackage{hyperref}                   
\usepackage{longtable}
\usepackage{natbib}                     
\usepackage{rotating}                   
\usepackage{setspace}                   
\usepackage{titlesec}                   
\usepackage{url}                        

\graphicspath{{./}}

\titleformat{\section}{\normalfont\sffamily\large\bfseries\color{black}}{\thesection.}{0.3em}{}
\titleformat{\subsection}{\normalfont\sffamily\small\bfseries\color{black}}{\thesubsection.}{0.3em}{}

\hypersetup{
	pdfauthor={\myname},
	pdftitle={\papertitle},
	pdfsubject={\papertitle},
	pdfkeywords={\paperkeywords},
	pdffitwindow=true,         
	breaklinks=true,           
	hidelinks=true             
}

\SetWatermarkText{DRAFT}
\SetWatermarkScale{1.3}
\SetWatermarkLightness{0.9}

\begin{document}

\title{\papertitle\footnote{{Preprint of: \papercitation}}}
\author[]{\myname\footnote{Email: \href{mailto:\myemail}{\myemail}}}
\affil[]{\myaffiliation}
\date{}

\maketitle

\begin{abstract}
As the rental housing market moves online, the Internet offers divergent possible futures: either the promise of more-equal access to information for previously marginalized homeseekers, or a reproduction of longstanding information inequalities. Biases in online listings' representativeness could impact different communities' access to housing search information, reinforcing traditional information segregation patterns through a digital divide. They could also circumscribe housing practitioners' and researchers' ability to draw broad market insights from listings to understand rental supply and affordability. This study examines millions of Craigslist rental listings across the US and finds that they spatially concentrate and over-represent whiter, wealthier, and better-educated communities. Other significant demographic differences exist in age, language, college enrollment, rent, poverty rate, and household size. Most cities' online housing markets are digitally segregated by race and class, and we discuss various implications for residential mobility, community legibility, gentrification, housing voucher utilization, and automated monitoring and analytics in the smart cities paradigm. While Craigslist contains valuable crowdsourced data to better understand affordability and available rental supply in real-time, it does not evenly represent all market segments. The Internet promises information democratization, and online listings can reduce housing search costs and increase choice sets. However, technology access/preferences and information channel segregation can concentrate such information-broadcasting benefits  in already-advantaged communities, reproducing traditional inequalities and reinforcing residential sorting and segregation dynamics. Technology platforms like Craigslist construct new institutions with the power to shape spatial economies, human interactions, and planners' ability to monitor and respond to urban challenges.
\vfill
\end{abstract}

\section{Introduction}

Large portions of the rental housing market have moved online over the past decade. Today, rental listings are primarily posted on websites like Craigslist, which holds a predominant position in the US as its 15\textsuperscript{th} overall most-visited website. According to the 2017 American Housing Survey, more renters in urbanized areas found their current homes through a site like Craigslist than through any other information channel. Housing practitioners and researchers, in turn, increasingly collect online listings to assess market supply in the smart cities paradigm of monitoring urban conditions through streams of user-generated data \citep{boeing_spot_2019,hu_monitoring_2019}.

Although online listings have recently become a primary mode of information exchange in US rental housing markets, little is known about how they function or how well they actually represent the full market \citep{boeing_new_2017,schachter_immigration_2017,besbris_language_2018}. If online listings are not representative---i.e., if sampling biases exist---what kinds of communities are over- or under-represented? Biases impact information equity and housing search costs as well as the conclusions that housing researchers and policymakers can draw about the real world from crowdsourced data \citep{mclaughlin_data_2018,arribas-bel_use_2018,folch_fast_2018,narayanan_toward_2019}. Yet little is currently known about spatial patterns or biases in the rental listings that compose these information landscapes and shape housing search outcomes.

This study assesses online rental market representativeness at the census tract scale, using a dataset of millions of Craigslist listings across the US. It explores the sociodemographics of over- and under-represented tracts and estimates spatial regression models to examine these traits' relationships with representation. It finds that listings are spatially concentrated and over-represent whiter, wealthier, and better-educated tracts. Majority-White tracts are over-represented more than twice as often as Hispanic or Black tracts. Although large swaths of this market are affordable to low-to-moderate income and Black families, homeseekers in whiter, wealthier, better-educated, and more-expensive communities have a surplus of information available online to aid their searches while seekers in other communities face a digital information deficit. The Internet helps democratize access to information but it does not necessarily equalize its supply. As the rental housing market moves online, technological self-selection and information supply biases construct new digital inequalities that shape housing search costs, choice sets, residential sorting, and the conclusions planners can draw about rapidly-evolving markets.

\section{Inequality in the Housing Information Landscape}

Housing technology platforms today forge emerging institutions with the capacity to reshape urban economies, human interactions, and information landscapes. \citep{wegmann_taming_2017,shaw_platform_2018,kim_planning_2019,fields_automated_2019,jiao_cities_2019}. These information landscapes---and any asymmetries or segregation within them---are central to housing search outcomes and residential sorting \citep{levitt_market_2008,kurlat_testing_2015,garmaise_confronting_2004,metzger_step_2019,ben-shahar_improved_2019}.

Prospective renters heterogeneously rely on a constellation of information sources to identify available units, including websites, newspapers, agents/brokers, property for-rent signs, and word of mouth. \citet{krysan_cycle_2017} theorize a two-stage search process in which homeseekers decide which neighborhoods to search and then choose which units within them to consider, emphasizing the importance of information supplies. \citet{rae_online_2015} contends that the Internet has emerged as the first port of call for such searches, but online information also impacts the second stage by rendering individual units and in turn neighborhoods more legible and accessible to seekers. As the Internet constitutes an ever-increasing share of the rental housing information supply, different communities' access to information and capacity to find housing depend both on listings' representativeness and communities' abilities/interests in engaging with these platforms. This can be considered from the supply-side (i.e., the supply of information by landlords, managers, and brokers) and the demand-side (i.e., homeseekers' Internet usage and search preferences).

On the supply-side, information about available housing units for rent traditionally appeared in local newspaper classifieds. Today, however, Craigslist has become the foremost such venue in the US, even as potential competitors like Facebook, Zillow, and Trulia try to challenge its near monopoly in online listings \citep{hau_newspaper_2006,brown_rental_2014,seamans_responses_2014,kroft_does_2014,yurieff_facebook_2017}. Researchers have increasingly turned to Craigslist listings to study individual metropolitan markets \citep[e.g.,][]{besbris_language_2018,brown_converting_2017,im_energy_2017,halket_homeownership_2015,mallach_meeting_2010,palm_scale_2018,schachter_immigration_2017,wegmann_understanding_2012}. Most studies of demographic representation on Craigslist have focused on discrimination by landlords \citep[e.g.,][]{hanson_field_2014,carlsson_discrimination_2014,evans_examining_2018,murchie_rental_2018} and the Fair Housing Act \citep[e.g.,][]{larkin_criminal_2010,oliveri_discriminatory_2010}. \citet{boeing_new_2017} examined Craigslist listings across the US, concluding further research was needed to understand sociodemographic submarket representativeness.

On the demand-side, who consumes this online supply of information about available rental units? Online housing search depends on Internet access and usage. The rise in Internet ubiquity over the past two decades has been accompanied by concerns about a digital divide between information \enquote{haves} and \enquote{have-nots} \citep{hersberger_are_2003,riddlesden_broadband_2014}. This divide may result from cultural differences or social inequalities: age, race, wealth, and education impact exposure and access to technology as well as attitudes, skills, and cultural norms in usage \citep{jones_u.s._2009,robinson_digital_2015}. Younger, whiter, better-educated, and higher-income Americans have higher Internet usage rates than other groups \citep{porter_using_2006}. Internet use and search engine behavior vary among different racial groups and as a function of native language \citep{slate_digital_2002,weber_who_2011}. Older adults are less likely to use the Internet, and this effect is more pronounced among those who are low-income, Black, or Hispanic \citep{choi_digital_2013}. The race gap is closing, however, and as of 2018, 89\% of White adults used the Internet versus 88\% of Hispanic and 87\% of Black adults, but usage remains lower today among older, less-educated, and lower-income Americans \citep{pew_research_center_internet_2018,pew_research_center_10_2019}.

It is less clear how this translates to housing search: different communities may prefer different information sources as a function of age, education, language, community ties, access to technology, and prior experiences with steering or discrimination. Pre-Internet, \citet{newburger_sources_1995} found that Black homebuyers relied on fewer information sources than Whites did, possibly due to housing information being harder to acquire in Black neighborhoods. \citet{farley_racial_1996} found that Blacks relied more on social ties and newspaper ads to find housing than Whites did, due to longstanding practices of real estate agents providing more information to White than Black homeseekers and steering them toward different neighborhoods. These agents also concentrate in Whiter and wealthier neighborhoods, focusing their services and benefits in already-advantaged communities \citep{besbris_investigating_2017}. Beyond race, \citet{deboer_resident_1985} argued that elderly homeseekers face higher search costs, more-constrained radii, and fewer information resources than younger seekers: these information asymmetries and inequalities produce housing market disequilibrium, with overpayment rates correlated with seeker characteristics \cite[cf.][]{desmond_are_2016,desmond_poor_2019}.

Do online housing markets attenuate these traditional information supply disparities or do they reproduce historical patterns of information segregation, steering, and sorting? As \citet[][p. 993]{stephens_gender_2013} argues, geospatial web platforms and technologies often \enquote{reproduce and exacerbate existing representational asymmetries} in the real world \cite[cf.][]{elwood_geographic_2010,mattern_interfacing_2014,thatcher_revisiting_2016,leszczynski_speculative_2016,brannon_datafied_2017}. Demographics and social networks shape our housing information supplies, producing neighborhood \enquote{blind-spots} and gaps in knowledge \citep{krysan_racial_2009}. Thus, information democratization and diversity on platforms like Craigslist could help expand and equalize residential searches. \citet{krysan_cycle_2017} argue that broadening such information sources---including rental search engine results---would expand homeseeker choice sets, in turn lessening the sociostructural factors that guide residential sorting \citep[cf.][]{sampson_neighborhood_2008,steil_household_2018}. While the Internet could offer information-disadvantaged homeseekers more information for their searches, it remains unclear if this potential has been realized \citep{palm_residential_2001,decker_housing_2010}. \citet{besbris_language_2018} found that the volume and type of information appearing in neighborhoods' Craigslist listings vary by demographics, with seekers in White- or Asian-majority neighborhoods receiving more information per listing than those in Black or Hispanic neighborhoods. \citet[][p.~598]{krysan_does_2008} found that Blacks were significantly less likely to use the Internet to search for housing than Whites, concluding: \enquote{Given the rapid growth of the Internet in renting and selling housing, the observed racial digital divide is a point of some concern.}

These studies of representation and participation mainly focus on \emph{demand}. The representativeness of available rental housing \emph{supply} online remains underexplored, yet it could impact equity in multiple ways. First, if Internet usage differs between demographic groups---and if certain kinds of communities appear more in online rental listings---neighborhood segregation could be perpetuated through information channel segregation and self-selection: forms of sociostructural steering. Second, it could reproduce longstanding housing information inequalities as new digital inequalities, where privileged communities obtain surpluses of relevant housing information---lessening their search costs---while others face relative deficits. Third, these information supplies could make certain units or neighborhoods more legible and accessible to homeseekers---with implications for gentrification, displacement, and housing voucher utilization.

\section{Methods}

\subsection{Data}

Due to its predominance in this information landscape today, Craigslist provides a powerful data source for studying the representativeness of online housing markets. This technology platform and the listings it contains shape human market behaviors while chronicling them for researchers. This study uses it to investigate if online rental listings over- or under-represent different communities and how community characteristics explain supply-side representativeness.

The study sites comprise the 12,505 census tracts within the core municipalities of the 50 most populous US metropolitan statistical areas, ignoring tracts that contain no rental units. We adopt the dataset of Craigslist rental housing listings collected in 2014 by \citet{boeing_new_2017}, which was filtered to remove duplicates and retain only geolocated listings, resulting in a set of 1.4 million listings (see \textit{ibid.} for a detailed explanation of the collection/cleaning process, along with summary statistics of each step).

Next we spatially join these tracts and rental listings then attach 2014 American Community Survey (ACS) tract-level data (Table \ref{tab:variables_list}). Although landlords, not neighborhoods, are the agents of interest, tract-level data offer the finest-scale approximation. The ACS is a sample survey, not an enumeration, and provides average rental vacancy rates rather than snapshots of vacancies at specific times.\endnote{See \citet{cresce_evaluation_2012}, \citet{spielman_patterns_2014}, and \citet{logan_models_2019} for comprehensive reviews and discussion of ACS tract-scale representativeness and limitations.} However, no existing dataset does so at sub-metropolitan scales. US Postal Service vacancy data provide finer spatial granularity, but crucially lack tenure information. Despite imperfections, the ACS offers the best data available to explore these questions by comparing average vacancy rates over time to Craigslist's aggregate listing volumes over time.

\subsection{Assessing Representation}

We tally how many listings appear in each tract as count $\kappa$. This represents the empirical/observed distribution. The ACS provides the number of vacant units for rent, $\tau$, in each tract. We then calculate a proportional reallocation, $\phi$, of these rental listings for each tract in each city as:

\begin{equation}
	\label{eq:allocation}
	\phi_t = \frac{\kappa_c \tau_t}{\tau_c}
\end{equation}

\begin{table}[tbp]
	\centering
	\small
	\caption{List of variables. Census sources refer to 2014 ACS tract-level data from which variable is derived. Percent estimates are converted to proportions by dividing by 100. \$ are 2014 inflation-adjusted US dollars.}
	\label{tab:variables_list}
	\begin{tabular}{l p{0.15\linewidth} p{0.675\linewidth}}
\toprule
Variable & Census Source & Description \\
\midrule
age2034  & DP05\_0008PE DP05\_0009PE & Proportion of population 20--34 years old\\
age65up  & DP05\_0021PE & Proportion of population 65 years and older\\
bb1940   & DP04\_0025PE & Proportion of structures built before 1940\\
black    & DP05\_0073PE & Proportion of population that is non-Hispanic black/African American\\
burden   & DP04\_0139PE DP04\_0140PE & Proportion of occupied rent-paying units paying gross rent > 30\% of household income\\
commute  & DP03\_0025E  & Mean travel time to work (minutes)\\
dcenter  &              & Straight-line distance (km) from tract centroid to city center\\
degree   & DP02\_0067PE & Proportion of population (25 years and older) with bachelor's degree or higher\\
density  & DP05\_0001E  & Total population (thousands) divided by land area (km\textsuperscript{2})\\
english  & DP02\_0111PE & Proportion of population (5 years and older) with English as only language spoken at home\\
foreign  & DP02\_0092PE & Proportion of population that is foreign-born\\
hhsize   & DP04\_0048E  & Average household size of renter-occupied units\\
hispanic & DP05\_0066PE & Proportion of population that is Hispanic/Latino\\
homeval  & DP04\_0088E  & Median value (\$, thousands) of owner-occupied housing units\\
income   & DP03\_0062E  & Median household income (\$, thousands)\\
male     & DP05\_0002PE & Proportion of population that is male\\
nonrels  & DP02\_0022PE & Proportion of household members that are non-relatives\\
poverty  & DP03\_0128PE & Proportion of families/people with income below poverty line\\
rent     & DP04\_0132E  & Median gross rent (\$, thousands) for occupied units paying rent\\
rooms    & DP04\_0036E  & Median rooms per housing unit\\
sameres  & DP02\_0079PE & Proportion of population (1 year and older) who lived in same home a year ago\\
singldet & DP04\_0007PE & Proportion of housing units that are single-unit detached\\
student  & DP02\_0057PE & Proportion of population (3 years and older) currently enrolled in college or graduate school\\
units    & DP04\_0046E DP04\_0005E & Count (thousands) of rental units\\
vacancy  & DP04\_0005E  & Ratio of vacant units for rent to total rental inventory\\
white    & DP05\_0072PE & Proportion of population that is non-Hispanic white\\
\bottomrule
\end{tabular}

\end{table}

where $\phi_t$ indicates how many of the observed Craigslist listings in city $c$ would appear in its tract $t$ if these listings were redistributed across $c$'s tracts according to each's proportion of $c$'s total vacant units for rent. This represents the theoretical/expected distribution. We then calculate\endnote{We add 1 to the numerator and denominator to avoid occasional division by zero and logarithm of zero.} each tract's over- or under-representation on Craigslist, $\lambda$, as:

\begin{equation}
	\label{eq:representation}
	\lambda_t = \frac{\kappa_t + 1}{\phi_t + 1}
\end{equation}

Thus, if $\lambda_t=1$, tract $t$ has the same number of rental listings on Craigslist that we would expect if the city's listings were redistributed among its tracts in proportion to each's share of the city's total vacant rental units. Higher and lower values indicate over- and under-representation respectively. Finally, we calculate Gini coefficients to measure how evenly the listings are distributed across city tracts. A coefficient of 1 indicates that a single tract contains all of the city's listings, while a coefficient of 0 indicates that they are perfectly evenly distributed among all of its tracts.

\subsection{Between-Group Differences}

Once we have assembled the ACS data and the Craigslist representation indicator $\lambda$, we calculate variables' differences-in-means between over-represented ($\lambda>1$) and under-represented ($\lambda<1$) tracts, as well as statistical significance via $t$-tests and practical significance (effect size) via Cohen's $d$:

\begin{equation}
	\label{eq:cohen_d}
	d = \frac{\mu_o - \mu_u}{\sigma_p}
\end{equation}

where $\mu_o$ and $\mu_u$ are the means of the over- and under-represented tracts and $\sigma_p$ represents their pooled standard deviation. This measures a standardized magnitude of difference between these two groups---namely, by how many standard deviations their means differ. By convention, a $d$ of 0.8 or greater represents a large effect, 0.5--0.8 represents a medium effect, 0.2--0.5 a small effect, and values below 0.2 a negligible effect \citep{cohen_power_1992}.

\subsection{Regression Analysis}

These differences can be unpacked while controlling for inventory, turnover, and other variables that influence listing volume. To investigate the \textit{ceteris paribus} associations between Craigslist representation and different sociodemographic and built environment characteristics, we estimate a multiple regression model (Model I) via ordinary least squares (OLS):

\begin{equation}
	\label{eq:regression_formula}
	y = \beta_0 + \beta_1 X_1 + \beta_2 X_2 + \epsilon
\end{equation}

where the response vector $y$ is tract Craigslist representation ($\lambda$), $\beta_0$ is the intercept, $X_1$ is a matrix of observations on tract sociodemographic and neighborhood variables, $X_2$ is a matrix of city dummy variables, $\epsilon$ is random error, and $\beta_1$ and $\beta_2$ are vectors of parameters to be estimated. To correctly specify a model that is linear-in-parameters, we log-transform the response\endnote{Ratios lack symmetry: when $\kappa < \phi$, $\lambda$ ranges from 0 to 1, but when  $\kappa > \phi$, $\lambda$ ranges from 1 to infinity. The logarithm corrects this: when $\kappa = \phi$, $\lambda = 1$ and $\log(\lambda) = 0$. It produces symmetry as $\log(a/b) = -\log(b/a)$. That is, $\log(a/b) = \log(a)-\log(b)$, so we evaluate the algebraic difference between logarithmic values.} and some of the predictors in $X_1$. Thus we can interpret the coefficients on log-transformed predictors as elasticities (the percent change in the response given a 1\% increase in the predictor) and those on untransformed predictors as semi-elasticities (the percent change in the response given a 1 unit increase in the predictor).

We control for intermetropolitan variation with $X_2$'s spatial fixed effects and for rental inventory and turnover with three variables in $X_1$: the count of rental units, the proportion of the population living in the same residence as a year ago, and the rental vacancy rate. Neighborhood character variables include the tract's median rooms per home, proportion of structures built before 1940, distance to the city center, and average commute time. These control for typical building size and age as well as location centrality and job accessibility.

Sociodemographic predictors include the tract's median household income, median gross rent, average renter household size, the proportions of the population 20--34 years old and 65 or older, the proportion currently enrolled in college/graduate school, the proportion with a bachelor's or graduate degree, the proportion that speaks English-only, and the proportions of the population that are White\endnote{\enquote{White} and \enquote{Black} are shorthand throughout for non-Hispanic White/Black.}, Black, or Hispanic. The model includes an interaction term---the White proportion $\times$ median income---to explore how race moderates the effect of income on representation.

Finally, Model I's diagnostics suggest the presence of spatial diffusion, so we estimate an additional spatial lag model (Model II) via maximum-likelihood estimation (MLE):

\begin{equation}
	\label{eq:regression_spatial_formula}
	y = \rho Wy + \beta_0 + \beta_1 X_1 + \beta_2 X_2 + \epsilon
\end{equation}

with variables defined as above, but also including $W$ as a queen-contiguity spatial weights matrix and $\rho$ as the spatial autoregressive coefficient to be estimated.

\section{Findings}

\subsection{Spatial Compression}

Across these cities, the tract-level observed distribution of Craigslist listings, $\kappa$, has a Gini coefficient of 0.80, while the expected distribution, $\phi$, has a coefficient of 0.70---i.e., rental listings are more spatially-concentrated than a proportional distribution would be---but this phenomenon is uneven between cities. In four markets (San Francisco, San Jose, Oklahoma City, and Las Vegas), the $\phi$ Gini is slightly higher (by 0.4\%--7.7\%) than that of $\kappa$, indicating rental listings are slightly more dispersed. However, in St. Louis, Providence, Miami, and Hartford, the $\kappa$ Gini is 70--120\% higher than that of $\phi$, indicating an extreme spatial compression of listings in these rental markets.

As this concentration suggests, most tracts are at least slightly under-represented on Craigslist (i.e., $\lambda$ < 1), but this varies by demographics. Overall, 52\% of majority-White tracts are over-represented, compared to only 19\% of majority-Black and 22\% of majority-Hispanic tracts. Examining the racial composition of tracts with $\lambda$ < 0.25 (i.e., with fewer than 25\% of the listings we expect), only 11\% of majority-White tracts are as such \enquote{very} under-represented, but 27\% of Black and 35\% of Hispanic tracts are.

\subsection{Differences Between Over- and Under-Represented Tracts}

\begin{figure}[tbp]
	\centering
	\includegraphics[width=1\textwidth]{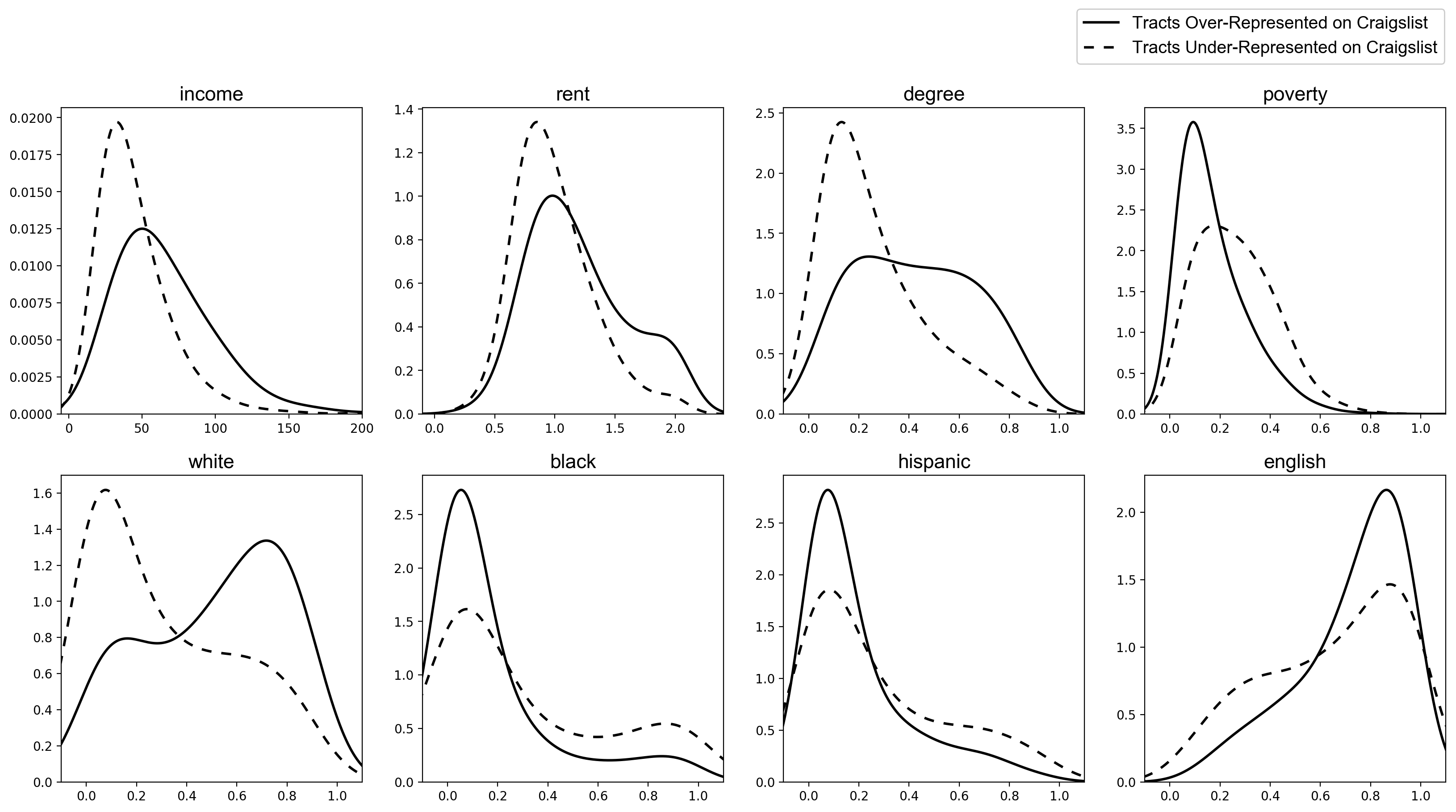}
	\caption{Variables' estimated probability densities across over/under- represented tracts: $x$-axis is in variable's units (details in Table \ref{tab:variables_list}). The probability for any interval equals the corresponding area under the curve.}
	\label{fig:variable_distributions}
\end{figure}

\begin{figure}[tbp]
	\centering
	\includegraphics[width=1\textwidth]{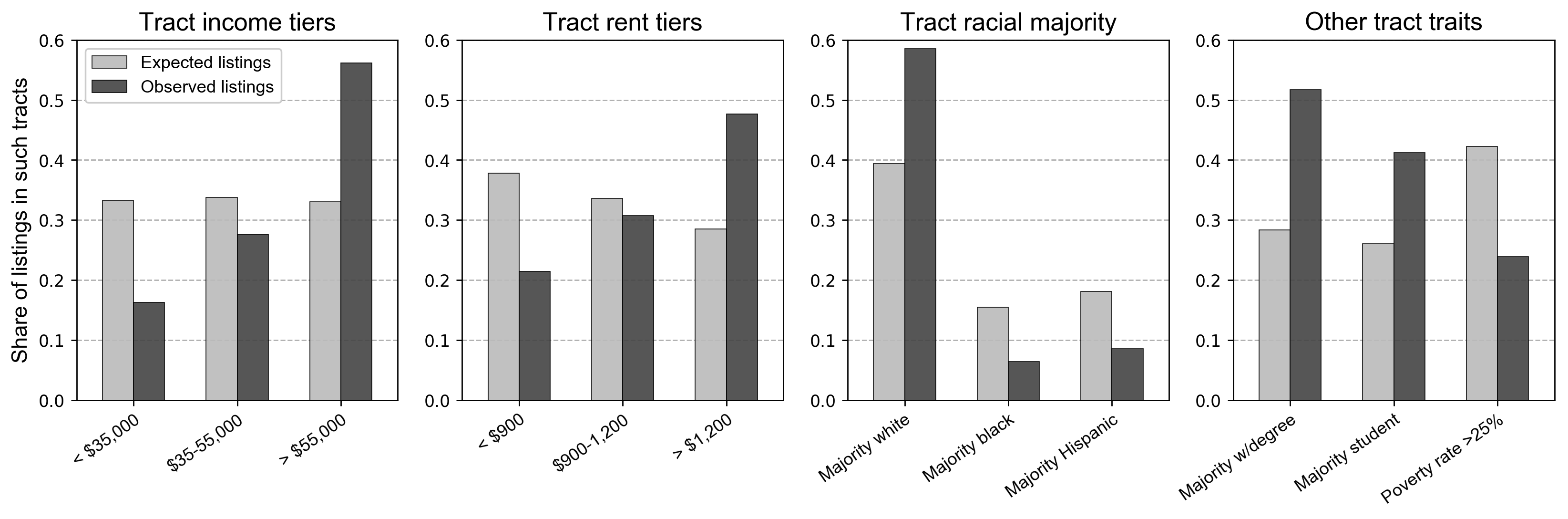}
	\caption{Share of total rental listings (expected versus observed distributions) in tracts with various characteristics.}
	\label{fig:tract_shares}
\end{figure}

Figures \ref{fig:variable_distributions} and \ref{fig:tract_shares} illustrate descriptive patterns in how sociodemographic and other neighborhood characteristics differ significantly between over- and under-represented tracts nationwide. Figure \ref{fig:variable_distributions} depicts the distributions of key variables, illustrating the differences between these two groups. In particular, the White proportion's distribution nearly inverts between over- and under-represented tracts. Six variables' differences demonstrate medium effect sizes, either positive or negative (Table \ref{tab:effects_over_under}\endnote{As a robustness test, we also recalculate $\delta$ in Table \ref{tab:effects_over_under} after discarding outliers (at $\pm2$ and $\pm3$ standard deviations) in each variable and $\lambda$. All values remain significant and signed in the same directions, with one exception: distance to city center.}): median income, gross rent, the White population proportion, the proportion with a bachelor's/graduate degree, the proportion enrolled in college/graduate school, the proportion below poverty, and the proportion experiencing rent burden. Twelve more have small, significant effects. Figure \ref{fig:tract_shares} shows how the expected and observed distributions of listings diverge. For instance, we expect only 33\% of all listings to appear in tracts with median income exceeding \$55,000, but instead we observe 56\%. Similarly we expect majority-White tracts to contain 39\% of all listings, but instead they contain 59\%.

\begin{table}[tb]
	\centering
	\small
	\caption{Differences-in-means between over/under-represented tracts nationwide: Cohen's $d$ represents effect size, $\delta$ represents difference, * indicates $t$-test significance at $p$ < 0.05.}
	\label{tab:effects_over_under}
	\begin{tabular}{l r r}
	\toprule
	{}       &   $d$ & $\delta$~~ \\ 
	\midrule
	degree   &  0.79 &     0.169* \\
	income   &  0.74 &    20.734* \\
	white    &  0.72 &     0.203* \\
	rent     &  0.58 &     0.206* \\
	student  &  0.39 &     0.074* \\
	homeval  &  0.35 &    79.116* \\
	singldet &  0.31 &     0.096* \\
	english  &  0.31 &     0.077* \\
	rooms    &  0.29 &     0.324* \\
	nonrels  &  0.23 &     0.014* \\
	age2034  &  0.21 &     0.021* \\
	male     &  0.16 &     0.007* \\
	age65up  &  0.07 &     0.004* \\
	dcenter  &  0.05 &     0.377* \\
	bb1940   & -0.15 &    -0.038* \\
	density  & -0.15 &    -1.348* \\
	sameres  & -0.19 &    -0.019* \\
	foreign  & -0.25 &    -0.043* \\
	hhsize   & -0.25 &    -0.185* \\
	hispanic & -0.33 &    -0.084* \\
	commute  & -0.44 &    -3.343* \\
	black    & -0.47 &    -0.141* \\
	burden   & -0.52 &    -0.076* \\
	poverty  & -0.61 &    -0.088* \\ 
	\bottomrule
\end{tabular}

\end{table}

\begin{table}[tbp]
	\centering
	\footnotesize
	\caption{Per-city difference-in-means effect sizes (Cohen's $d$) between over/under-represented tracts: * indicates corresponding difference-in-means $t$-test significance at $p$ < 0.05.}
	\label{tab:effects_cities}
	\fontsize{9}{10}\selectfont
\begin{tabular}{lrrrrrrrr}
	
	\toprule
	{}                 & income &    rent &  degree & poverty & student & english &   white &  hhsize \\ \midrule
	Atlanta, GA        &  0.45* &   0.73* &   0.69* &  -0.45* &   0.42* & -0.32~~ &   0.69* &  -0.49* \\
	Austin, TX         &  0.63* &   0.40* &   0.78* &  -0.50* &  0.08~~ &   0.64* &   0.76* & -0.19~~ \\
	Baltimore, MD      &  0.79* &   0.62* &   1.21* &  -0.53* &   0.96* &  -0.50* &   1.03* &  -0.61* \\
	Birmingham, AL     &  0.84* &   0.71* &   0.81* &  -0.88* &  0.24~~ & -0.14~~ &   0.61* &  0.00~~ \\
	Boston, MA         &  0.67* &   0.60* &   0.98* &  -0.36* &   0.60* &   0.57* &   1.03* &  -0.98* \\
	Buffalo, NY        &  0.50* &   0.61* &  0.11~~ & -0.21~~ &  0.19~~ &  0.26~~ &  0.30~~ &  0.32~~ \\
	Charlotte, NC      &  0.63* &   0.60* &   0.74* &  -0.64* &   0.34* &   0.30* &   0.58* & -0.16~~ \\
	Chicago, IL        &  1.51* &   1.33* &   1.62* &  -0.89* &   1.04* &   0.32* &   1.36* &  -0.82* \\
	Cincinnati, OH     &  0.54* &   0.82* &   0.62* &  -0.45* &   0.46* &  -0.53* &   0.84* & -0.01~~ \\
	Cleveland, OH      &  0.51* &  0.32~~ &   0.59* &  -0.37* &   0.40* &  -0.49* &   0.63* & -0.17~~ \\
	Columbus, OH       &  0.92* &   0.83* &   0.72* &  -0.79* & -0.03~~ &   0.38* &   0.72* & -0.18~~ \\
	Dallas, TX         &  0.90* &   0.91* &   1.01* &  -0.83* &   0.67* &   0.61* &   0.97* &  -0.49* \\
	Denver, CO         &  0.51* &   0.65* &   0.42* &  -0.47* &  0.27~~ &   0.40* &   0.41* & -0.12~~ \\
	Detroit, MI        & 0.22~~ &   0.42* &   0.34* & -0.26~~ &  0.30~~ &  0.22~~ &  0.06~~ &  0.08~~ \\
	Hartford, CT       & 0.57~~ &  0.28~~ &   1.90* & -0.59~~ &  0.79~~ &  0.54~~ &  0.93~~ &  -1.52* \\
	Houston, TX        &  0.99* &   0.95* &   0.99* &  -0.98* &   0.50* &   0.73* &   1.04* &  -0.44* \\
	Indianapolis, IN   &  0.69* &   0.42* &   0.60* &  -0.69* &  0.11~~ &   0.30* &   0.68* & -0.22~~ \\
	Jacksonville, FL   &  0.59* &   0.39* &   0.57* &  -0.44* &  0.04~~ & -0.13~~ &   0.43* & -0.08~~ \\
	Kansas City, MO    &  0.62* &   0.72* &   0.57* &  -0.51* &  0.16~~ &  0.28~~ &   0.45* & -0.03~~ \\
	Las Vegas, NV      &  1.01* &   1.13* &   0.69* &  -0.81* &  0.18~~ &   0.73* &   0.65* &   0.39* \\
	Los Angeles, CA    &  0.80* &   0.88* &   0.78* &  -0.67* &   0.40* &   0.65* &   0.79* &  -0.35* \\
	Louisville, KY     &  1.15* &   0.79* &   0.88* &  -0.88* &   0.46* &  0.14~~ &   0.86* & -0.31~~ \\
	Memphis, TN        &  0.89* &   0.89* &   0.63* &  -0.82* &  0.14~~ & -0.08~~ &   0.61* & -0.03~~ \\
	Miami, FL          &  1.68* &   1.54* &   2.08* &  -1.37* &   1.26* &  0.17~~ &   2.00* &  -1.18* \\
	Milwaukee, WI      &  0.77* &   0.56* &   0.99* &  -0.59* &   0.98* &  0.29~~ &   0.90* &  -0.47* \\
	Minneapolis, MN    & 0.35~~ &  0.32~~ &   0.67* & -0.31~~ &  0.34~~ &   0.40* &   0.59* & -0.35~~ \\
	Nashville, TN      &  0.70* &   0.98* &   1.00* &  -0.68* &   0.35* &  0.01~~ &   0.79* & -0.25~~ \\
	New Orleans, LA    &  1.14* &   1.11* &   1.51* &  -0.90* &   0.86* & -0.28~~ &   1.59* &  -0.36* \\
	New York, NY       &  0.68* &   0.67* &   0.97* &  -0.36* &   0.59* &   0.37* &   0.52* &  -0.62* \\
	Oklahoma City, OK  &  0.74* &   0.46* &   0.72* &  -0.59* &  0.26~~ &   0.51* &   0.63* & -0.07~~ \\
	Orlando, FL        &  0.62* &   0.69* &   0.65* &  -0.54* &  0.27~~ &  0.05~~ &   0.56* & -0.26~~ \\
	Philadelphia, PA   &  0.75* &   0.76* &   1.19* &  -0.38* &   1.01* &   0.31* &   0.70* &  -0.61* \\
	Phoenix, AZ        &  0.74* &   0.66* &   0.60* &  -0.81* &   0.21* &   0.66* &   0.73* & -0.01~~ \\
	Pittsburgh, PA     &  0.75* &   0.58* &   0.76* &  -0.72* &  0.27~~ & -0.26~~ &   0.83* & -0.11~~ \\
	Portland, OR       &  0.47* &   0.44* &   0.48* &  -0.45* & -0.06~~ &   0.43* &   0.46* & -0.28~~ \\
	Providence, RI     & 0.18~~ & -0.02~~ &  0.26~~ &  0.09~~ &  0.13~~ &  0.62~~ &  0.38~~ & -0.48~~ \\
	Raleigh, NC        &  0.67* &  0.14~~ &   0.89* &  -0.46* &  0.09~~ &  0.34~~ &   0.67* & -0.10~~ \\
	Richmond, VA       & 0.37~~ &   1.12* &  0.51~~ & -0.05~~ &  0.50~~ &  -0.64* &  0.49~~ &  0.13~~ \\
	Riverside, CA      & 0.44~~ &   0.70* &  0.38~~ &  -0.50* & -0.05~~ &  0.20~~ &  0.33~~ & -0.27~~ \\
	Sacramento, CA     &  0.75* &   0.57* &   0.65* &  -0.72* &  0.04~~ &   0.39* &  0.27~~ & -0.35~~ \\
	Salt Lake City, UT & 0.46~~ &  0.51~~ &  0.00~~ & -0.25~~ & -0.26~~ & -0.14~~ & -0.14~~ &  0.48~~ \\
	San Antonio, TX    &  1.01* &   0.84* &   0.72* &  -0.71* &  0.04~~ &   0.87* &   0.80* &  0.05~~ \\
	San Diego, CA      &  1.02* &   1.04* &   0.72* &  -0.81* & -0.12~~ &   0.40* &   0.43* &  0.06~~ \\
	San Francisco, CA  &  0.41* &   0.50* &  0.24~~ &  -0.41* & -0.03~~ &  0.14~~ &  0.10~~ &  0.12~~ \\
	San Jose, CA       &  0.59* &   0.54* &   0.31* &  -0.51* & -0.13~~ &  0.16~~ &  0.13~~ &  0.25~~ \\
	Seattle, WA        & 0.24~~ &  0.19~~ &  0.27~~ & -0.04~~ &  0.24~~ &  0.12~~ &  0.22~~ & -0.02~~ \\
	St. Louis, MO      &  0.70* &  0.39~~ &   0.79* &  -0.48* &   0.62* & -0.36~~ &   0.77* &  -0.56* \\
	Tampa, FL          &  0.77* &   0.63* &   0.59* &  -0.67* & -0.04~~ &  0.36~~ &   0.45* & -0.14~~ \\
	Virginia Beach, VA & 0.40~~ &   0.59* & -0.04~~ & -0.23~~ & -0.12~~ & -0.14~~ &  0.24~~ &  0.39~~ \\
	Washington, DC     &  1.06* &   1.23* &   1.34* &  -0.91* &   0.82* &  -0.77* &   1.15* &  -0.52* \\ \bottomrule
\end{tabular}
\normalsize
\end{table}

On average compared to under-represented tracts, over-represented tracts have a White population proportion 20 percentage-points\endnote{We distinguish between percentage-point differences, which are additive, and percent changes, which are not.} (pp) higher, a Hispanic proportion 8 pp lower, and a Black proportion 14 pp lower (Table \ref{tab:effects_over_under}). The proportion that speaks English-only is 8 pp higher and the foreign-born proportion is 4 pp lower. Average median home values are nearly \$80,000 higher, median incomes \$21,000 higher, and median gross rents \$200 higher. The proportion of the population with a bachelor's/graduate degree is 17 pp higher and the proportion currently enrolled in college/graduate school is 7 pp higher, on average. In over-represented tracts the proportion of the population below poverty is 9 pp lower and the rent-burdened proportion is 8 pp lower. Over-represented tracts offer shorter commutes and have slightly lower population densities, smaller household sizes, and higher proportions of single-unit detached housing on average.

These differences are significant, but are they consistent across cities? Although Craigslist represents certain cities more evenly than others, some consistent patterns emerge in the per-city differences between over- and under-represented tracts (Table \ref{tab:effects_cities}). Average median income is higher in over-represented tracts (i.e., $d$ > 0) in every city, and more than 1.5 standard deviations higher in Miami and Chicago. Rents are higher in over-represented tracts in every city with a significant difference, and more than 1 standard deviation higher in 7 cities, again led by Miami and Chicago. The proportion of the population with a bachelor's/graduate degree is higher in every city with a significant difference, and more than 1.5 standard deviations higher in Miami, Hartford, Chicago, and New Orleans. The proportion enrolled in college/graduate school is higher in every city with a significant difference, and more than 1 standard deviation higher in Miami, Chicago, and Philadelphia. Similarly, the White proportion is higher in every city with a significant difference, and more than 1 standard deviation higher in 7 of these, led by Miami and New Orleans. However, the proportion that speaks English-only is more divisive: while some cities have medium-sized negative effects, many more demonstrate medium or large positive effects. Finally, the proportion below poverty is lower in every city with a significant difference, led by Miami, Houston, Washington, and New Orleans, and the average renter household size is lower in all but one city (Las Vegas) with a significant difference, led by Hartford, Miami, and Boston.

\subsection{Regression Results}

\begin{table}[tb]
	\centering
	\small
	\caption{Regression model parameter estimates and standard errors. Variables/units defined in Table \ref{tab:variables_list}. Model I (Equation \ref{eq:regression_formula}) is estimated with OLS. Model II (Equation \ref{eq:regression_spatial_formula}) is estimated with MLE. Spatial fixed effects not reported. Dependent variable is Craigslist representation ($\lambda$) and * indicates significance at $p$ < 0.05.}
	\label{tab:regression_results}
	\begin{tabular}{l r r r r}
	\toprule
	                         & \textbf{Model I} &       & \textbf{Model II} &       \\ 
	\midrule
	                         & Estimate         & SE    & Estimate          & SE    \\ 
    \midrule
	constant                 & -0.730*          & 0.316 & -0.425~~          & 0.313 \\
	spatial lag ($\rho$)     & ---~~            & ---   & 0.582*            & 0.032 \\
	units                    & -0.098*          & 0.020 & -0.106*           & 0.020 \\
	vacancy                  & -11.476*         & 0.165 & -11.326*          & 0.163 \\
	sameres                  & -0.399*          & 0.147 & -0.385*           & 0.145 \\
	dcenter\_log             & -0.118*          & 0.020 & -0.134*           & 0.019 \\
	commute\_log             & -0.069~~         & 0.070 & -0.028~~          & 0.069 \\
	bb1940                   & -0.283*          & 0.057 & -0.246*           & 0.056 \\
	rooms                    & 0.044*           & 0.019 & 0.051*            & 0.019 \\
	rent                     & 0.575*           & 0.052 & 0.500*            & 0.052 \\
	income\_log              & 0.359*           & 0.048 & 0.350*            & 0.047 \\
	age2034                  & 0.406*           & 0.196 & 0.294~~           & 0.193 \\
	age65up                  & -0.498*          & 0.223 & -0.461*           & 0.220 \\
	student                  & 0.140~~          & 0.094 & 0.161~~           & 0.093 \\
	english                  & 0.686*           & 0.121 & 0.571*            & 0.120 \\
	hhsize\_log              & -0.212*          & 0.064 & -0.224*           & 0.064 \\
	degree                   & 1.034*           & 0.112 & 0.867*            & 0.110 \\
	white                    & 0.734*           & 0.350 & 0.787*            & 0.345 \\
	black                    & -0.572*          & 0.137 & -0.424*           & 0.136 \\
	hispanic                 & -0.296*          & 0.106 & -0.218*           & 0.105 \\
	white$\times$income\_log & -0.333*          & 0.081 & -0.330*           & 0.080 \\ 
	\midrule
	$R^2$                    & 0.46~~           &       & ---~~             &       \\
	Pseudo-$R^2$             & ---~~            &       & 0.47~~            &       \\
	AIC                      & 35895~~          &       & 35685~~           &       \\
	Log-Likelihood           & -17879~~         &       & -17773~~          &       \\ 
	\bottomrule
\end{tabular}
\end{table}

As the two regression models tell similar stories (Table \ref{tab:regression_results}), we focus on Model I for interpretability.\endnote{Due to spatial diffusion, Model II's coefficients cannot be directly interpreted as marginal effects because the response depends on the predictors' values at all spatially correlated locations \citep[][p. 164]{anselin_modern_2014}.} A \textit{ceteris paribus} \$10 increase in tract median rent increases representation on Craigslist by 0.58\%, a 1 pp increase in the proportion with a bachelor's/graduate degree increases it 1.03\%, and a 1 pp increase in the proportion that speaks English-only increases it 0.69\%. A 1 pp increase in the 20--34 year old population proportion increases it by 0.41\% but a 1 pp increase in the proportion age 65 and older decreases it 0.50\%. A 1\% increase in distance from the city center decreases tract representation on Craigslist by 0.12\% and a 1 pp increase in the proportion of structures built before 1940 decreases it 0.28\%. Although a 1 room increase per home increases representation by 4.4\%, a 1\% increase in renter household size decreases it 0.21\%.

A 1 pp increase in the Black population proportion decreases representation by 0.57\% and a 1 pp increase in the Hispanic proportion decreases it 0.30\%. Thus, even when controlling for income, education, age, and language, greater shares of Black or Hispanic residents predict lower representation in online rental listings, indicating a distinct race/ethnicity effect. Given the interaction term, the marginal effect of income on Craigslist representation depends on the White population proportion. When the population is 10\% White, a 1\% increase in median income increases representation by 0.33\%. But with a 90\% White population, a 1\% increase in median income increases it by only 0.06\%. Thus, higher incomes are associated with greater online representation in less-White communities, but homogeneously White communities tend to be represented similarly online nearly irrespective of income. Likewise, in poorer tracts, \enquote{whiteness} has a more positive effect: when median income is \$8,000, a 1 pp increase in the White population proportion increases Craigslist representation by 0.04\%. But when median income is \$100,000, a 1 pp increase in the White proportion decreases it by 0.80\%. In other words, having a larger proportion of White residents predicts greater online representation in very poor communities, but in high-income communities it predicts lower representation: the model's interaction/moderation effects suggest that wealthier Whites may be more likely to hire brokers or rely on more-exclusive listing services than the free-and-open Craigslist platform.

Finally, we perform a series of robustness tests,\endnote{These tests include re-estimating Equations \ref{eq:regression_formula} and \ref{eq:regression_spatial_formula} after: removing response variable outliers (at $\pm2$ and $\pm3$ standard deviations); operationalizing a location quotient as an alternative response variable \citep[see][]{billings_location_2012}; filtering the tracts by various minimum-listings thresholds; adding a spatial-error term to each equation \citep[see][]{anselin_modern_2014}; and computing an alternative distance-decay spatial weights matrix.} across which we find the models tell a consistent story, especially regarding key variables of interest. For example, the proportion-Black and proportion-Hispanic parameter estimates remain negative across these tests and remain significant across nearly all of them. The median rent and median income parameter estimates remain positive and significant across all the robustness tests. In other words, these key race/ethnicity and socioeconomic status trends persist throughout various analytical reformulations.

\section{Discussion}

\subsection{Representativeness}

These data distill the collective behavior and intentions of millions of people listing rental units online. They do not, however, represent all market segments and activity. It becomes essential to understand these biases as homeseekers and researchers increasingly turn to the Internet to explore rental markets. Craigslist listings spatially concentrate more than expected and some markets---such as Miami's---demonstrate extreme compression. Such cities consistently appear among those with the largest differences between over- and under-represented tracts across multiple sociodemographic variables. Their online housing markets are spatially concentrated and digitally segregated by race and class.

Miami's disparities starkly illustrate this. Its average population proportion with a college degree and White proportion are both 2 standard deviations higher in over- versus under-represented tracts. Its average median income and rent are both 1.5 standard deviations higher. On average in Miami, 51\% of the population has a degree in over-represented tracts versus only 17\% in under-represented tracts. 33\% versus 7\% of the population is White, household incomes are \$67,000 versus \$28,000, and home values are \$371,000 versus \$163,000. Nationwide, tracts over-represented on Craigslist are significantly better-educated, whiter, and wealthier than under-represented tracts. They have higher rents and larger homes, but smaller household sizes. They contain more college students, more English-only speakers, and fewer immigrants. Majority-White tracts are over-represented 2.4 and 2.7 times as often as Hispanic and Black tracts. Conversely, only 1 in 9 majority-White tracts is \enquote{very} under-represented, compared to over a quarter of Black and over a third of Hispanic tracts. Craigslist's information-broadcasting benefits concentrate in traditionally advantaged communities.

Controlling for confounds to disentangle relationships, we find that higher rents and larger proportions of the population that are college-educated, aged 20--34, and speak English-only each significantly predict greater representation on Craigslist. Newer housing stock and more rooms per home---but smaller renter household sizes---also predict greater representation. Higher incomes have a universally positive relationship with representation, but a larger one in tracts with higher minority populations. Finally, the key race/ethnicity effects persist in both models and throughout the robustness tests: larger Black and Hispanic shares of the population consistently predict lower representation in online rental listings.

\subsection{Digital Inequality or a Redoubt of Illegibility?}

Beyond identifying sampling biases that render certain communities less-legible to big data analytics, these findings also contribute supply-side evidence to theories of the digital divide and disparate housing search. For instance, the literature suggests that Black and elderly Americans are less likely to search for housing online, and we confirm that these communities are less-represented in online listings. Our findings on immigrants and non-English speakers may relate to common preferences to use soft ties rather than English-language websites to advertise or seek housing. Craigslist's information channel dominance varies accordingly within and between cities.

Of course, certain communities may prefer certain information channels for a variety of reasons. But as the housing market moves online, technological self-selection among listers and seekers reinforces the sociostructural sorting mechanisms that perpetuate neighborhood blind-spots, residential segregation, and unequal search costs. This study identifies structural differences in this information supply between different kinds of communities, but not \emph{why} each submarket behaves differently---many factors could be simultaneously at play. Yet it demonstrates that the spatial patterns of traditional housing information inequality and steering are reproduced online through the segregation and asymmetry of information supplies.

These phenomena have various implications for residential mobility, sorting, and housing policy. As homeseekers tend to search for units in neighborhoods they are already familiar with \citep{krysan_cycle_2017}, disadvantaged communities miss out on the reduced search costs and expanded choice sets provided by online platforms---benefits that instead primarily accrue to already-advantaged communities. A two-tiered system emerges in which privileged communities exchange housing information through one channel, while all others resort to separate channels. These forces perpetuate the self-reinforcing cycles of durable inequality: information segregation limits homeseekers' discovery of housing in neighborhoods different from their own---in turn limiting the ability to integrate neighborhoods for more diversity of incomes, education levels, ages, and ethnicities \cite[cf.][]{krysan_cycle_2017,ellen_can_2018,pendall_pathways_2018,steil_household_2018}.

Conversely, another important implication is that \emph{not} appearing in online listings could be advantageous for underprivileged neighborhoods if it makes them less accessible to gentrification and displacement. That is, if these communities' members tend to search for housing online less than wealthier seekers do, and if available units in these neighborhoods appear online less, they may be less discoverable to Internet-savvy outsiders. Thus, information under-representation could shield vulnerable communities in a redoubt of digital illegibility, while reciprocally making over-represented privileged communities more legible to a broad cross-section of homeseekers who otherwise may not have acquired information about these neighborhoods. The latter has important policy ramifications as it could expand the choice sets of housing voucher holders seeking to move to higher-opportunity neighborhoods, if affordable units can be found \cite[cf.][]{rosen_rigging_2014,schwartz_encouraging_2017,mclaughlin_data_2018,reina_section_2019}. Practitioners, for example, could take advantage of this by broadcasting eligible-unit information online or by helping voucher holders devise online search strategies that focus on unit characteristics in higher-opportunity neighborhoods rather than familiar geographic silos.

Finally, what about market segmentation? Concerns about digital inequality could be assuaged if online listings match high-end housing supply and demand, while other segments simply use a different suitable information channel instead. But, crucially, Craigslist does not merely represent the high-end of the market. 49\% of these listings are affordable (i.e., asking rent < 30\% of income) for households just below the 2014 US median household income, and 23\% of them are affordable for households just below the corresponding Black median. Millions of Craigslist rental listings \emph{are} within the reach of low-to-moderate income and Black families---yet listings appear disproportionately infrequently in these communities. Housing brokers historically provided different information to White and Black homeseekers: although the Internet promises information democratization, these disparities seem to continue online.

\subsection{Information Landscapes and Housing Dynamics}

A critical project lies ahead in unpacking the cause-and-effect/supply-and-demand question. Do landlords in certain communities advertise online less because they know their market segment uses the Internet less to search for housing? Or do certain communities use the Internet less to search for housing because of unfamiliarity, its dearth of relevant information, or past experiences with racialized interactions? If a community member rarely finds suitable units when she does search online, why would she continue doing so? Conversely, a relative affinity for Internet use in wealthier, whiter, better-educated neighborhoods could stimulate the local supply of online information. Local landlords in certain neighborhoods may prefer to advertise online as their own Internet usage reflects that of the surrounding community. Neighborhoods with more subsidized housing might rely on other information channels, even other online channels such as https://gosection8.com/. Some landlords even avoid Craigslist to deliberately filter and limit the number of potential applicants \citep{mendez_professional_2016}. Such patterns reflect path-dependent interactions of supply and demand feedback loops, suppressing some communities' usage while buttressing others'.

Although Craigslist does not constitute the entire online listings marketplace, it holds a dominant market position and has uniquely low barriers to entry. However, these listings were collected during spring and could reflect seasonal biases: future work can confirm these findings with longer-term longitudinal data and larger sample sizes to lessen the lumpiness of real-world volunteered geographic information. This study examined large core cities, not wider metropolitan areas or small towns. Although many core cities contain extensive suburban neighborhoods, the inclusion of further-flung suburbs and exurbs might change estimated model parameters: future research can examine regional markets or estimate individual local models. Finally, due to the discussed limitations of ACS vacant unit counts, follow-up work should confirm these results with additional sources of vacancy data.

Our findings provide a first glimpse into this information landscape. This study modeled tract-level representation, but subsequent research can investigate how demographic, neighborhood, and unit characteristics contribute to asking rents and the difference between asking rents and corresponding ACS rents. This leads to a critical open question: do online listings render units in disadvantaged neighborhoods more accessible to privileged homeseekers, or does under-representation shield these communities from \enquote{colonization?} Given the substantial demographic disparities identified here, future work can explore how tracts change over time to investigate online listings' role in housing search efficiency and the broader dynamics of gentrification.

\section{Conclusion}

If information is power in today's cities and markets, then disparate access to it can influence community futures, opportunities, and equity. As online platforms capture an ever-greater share of the rental housing market's information landscape, data sources like Craigslist offer invaluable opportunities to understand affordability and available rental supply in real-time. They do not, however, provide a holistic view of the market, as this information exchange over- and under-represents certain communities.

This study developed a methodology to identify sampling biases in large harvested urban datasets, providing a glimpse into the supply-side of digital inequality and the potential blind-spots of smart city automated analytics. The Internet's ability to broaden information sources and democratize access to housing data remains contingent on landlords listing online. Housing seekers in whiter, wealthier, better-educated, and more-expensive communities have a surplus of relevant information online to aid their searches while seekers in other communities face a deficit, reproducing historical patterns of information inequality. In turn, these biases influence the reinforcing feedback loops of supply and demand in community usage and residential sorting, as well as the conclusions that housing planners can draw by collecting these data.

\section*{Acknowledgments}

Portions of this work were previously presented at the ACSP annual conference and at Harvard University’s Joint Center for Housing Studies. The author wishes to thank those attendees as well as Max Besbris, Ariela Schachter, Jake Wegmann, Nat Decker, Sam Maurer, Craig Jones, Dan Chatman, and Dillon Mahmoudi for their helpful comments and suggestions. Finally, this work is particularly indebted to Paul Waddell, who graciously allowed the reuse of data originally collected in his lab at UC Berkeley.

\IfFileExists{\jobname.ent}{\theendnotes}{}

\setlength{\bibsep}{0.00cm plus 0.05cm} 
\bibliographystyle{SageH}

\end{document}